\documentclass[final]{svjour2}
\usepackage{graphicx}
\usepackage{rotating}
\usepackage{amssymb}
\usepackage{mathptmx}
\usepackage[numbers]{natbib}
\makeatletter
\journalname{Journal of Low Temperature Physics}

\bibpunct{}{}{,}{s}{}{,}

\begin{document}

\newcommand{\hdblarrow}{H\makebox[0.9ex][l]{$\downdownarrows$}-}
\title{Multi-Chroic Feed-Horn Coupled TES Polarimeters}

\author{J. McMahon$^1$ \and J. Beall$^2$ \and  D. Becker$^{2,3}$ \and  H.M. Cho$^2$, R. Datta$^1$ \and  A. Fox$^{2,3}$ \and  N. Halverson$^3$ \and  J. Hubmayr$^{2,3}	$ \and  K. Irwin$^2$ \and   J. Nibarger$^{2,3}$ \and  M. Niemack$^2$ \and  H. Smith$^1$}

\institute{1:Department of Physics, University of Michigan, Ann Arbor, MI 48109, USA\\
\email{jeffmcm@umich.edu}
\\2:National Institute of Standards and Technology,  Boulder, CO 80305, USA
\\3: University of Colorado, Boulder, CO 80309
\\4:Department of Astrophysical Sciences, University of Colorado, Boulder, CO,    80309 , USA\\}

\date{07.31.2011}

\maketitle


\begin{abstract}
Multi-chroic polarization sensitive detectors offer an avenue to increase both the spectral coverage and sensitivity of instruments optimized for observations of the cosmic-microwave background (CMB) or sub-mm sky. We report on an effort to adapt the Truce Collaboration horn coupled bolometric polarimeters for operation over octave bandwidth.  Development is focused on detectors operating in both the 90 and 150 GHz bands which offer the highest CMB polarization to foreground ratio. We plan to deploy an array of 256 multi-chroic 90/150 GHz polarimeters with 1024 TES detectors on ACTPol in 2013, and there are proposals to use this technology for balloon-borne instruments. The combination of excellent control of beam systematics and sensitivity make this technology ideal for future ground, ballon, and space missions.

\end{abstract}

\section{Introduction}
Measurements of the CMB are progressing rapidly\cite{reichardt09,fowler:2010cy,Das:2010ga,chiang09,piacentini06,montroy06,bischoff08,sievers07,leitch05,Wu:2006ji,brown09,2010arXiv1012.3191Q,Lueker:2009rx,Keisler:2011aw,nolta09} yet many faint signals, containing rich cosmological information, remain unmeasured.  Sensitive observations of CMB polarization will provide information about the energy scale of inflation (from the large angular scale power spectrum)  as well as tight measurements of neutrino masses and early dark energy (through gravitational lensing).  Recovering these signals requires instruments with high sensitivity, excellent control of systematics, and multi-wavelength coverage to deal with foregrounds.
 
Sensitivity and spectral coverage can be improved by detecting multiple frequency bands in each pixel (multi-chroic detectors).  In \S \ref{sec_multichroix} we describe our efforts to extend the Truce feed horn-coupled polarimeter to accommodate multiple wide bands.   This architecture offers excellent control of beam systematic effects and polarization axes that are independent of frequency.  In \S \ref{sec:act} we describe plans to field an array of these detectors as part of the ACTPol experiment.   We conclude in \S \ref{sec:conc}.
\section{Multi-Chroic Detectors}
  \label{sec_multichroix}

\begin{figure}[t!]
\begin{center}
\resizebox{0.9\hsize}{!}{
  \includegraphics[width=8in]{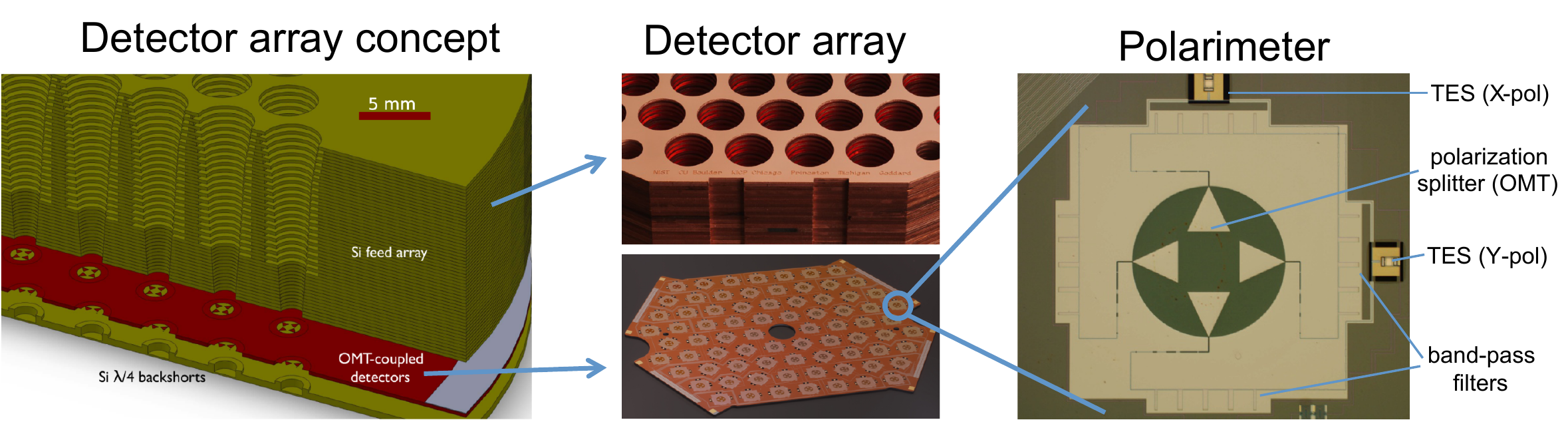}}
\end{center}
  \caption{\textit{Left:} The assembly of a single frequency Truce detector array. The
    array will consist of a silicon micro-machined platelet array of
    corrugated feedhorns above a detector wafer and a micro-machined
    backshort wafer. The assembly of feedhorns, detector array and
    the backshort wafer are separated in this diagram for
    clarity. \textit{Center:} Photograph of a prototype horn array (top) and prototype detector wafer (bottom). \textit{Right:} Zoomed in photograph of a single
    detector in an array.  The major components are labeled including the planar
    OMT, the stub filters, and the TES islands.
  \label{fig:probes}}
\end{figure}

The multi-chroic arrays under development are an extension of the single-frequency Truce detectors \cite{yoonltd13}.  (See Figure \ref{fig:probes}) The optical beam is defined by a corrugated feedhorn, which has a long history of use in polarimeters in radio astronomy and CMB observations. Corrugated horns offer low sidelobes, circularly symmetric beams and a low cross-polarization over a wide bandwidth --- ideal characteristics for use in polarimetry. The waveguide output of the horn couples to a planar ortho-mode transducer (OMT) \cite{mcmahon09} which separates the orthogonal polarization components of the incoming radiation into independent co-planar waveguides (CPW).  The OMT defines a polarization axis that is independent of frequency.   Each CPW transitions to a microstrip (MS) transmission line using a numerically optimized stepped impedance transformer.  The signals then pass through a set of resonant MS stub filters which define the spectral pass-band of the detectors.

Once filtered, the incident power for each polarization is dissipated on a suspended and thermally isolated silicon nitride (Si$_{3}$N$_{4}$) island using lossy gold MS terminations, and the corresponding temperature change of this `bolometer island' is detected by a MoCu transition edge sensor (TES). Summing the outputs of the two bolometers yields the temperature of the incoming radiation, and differencing the two yields the Stokes Q parameter. The full linear polarization state of a given location on the sky can be measured by combining measurements of multiple polarimeters rotated relative to each other, or by observing the sky at multiple parallactic angles.


%
 \begin{figure}[t!]
\includegraphics[width=4.5in]{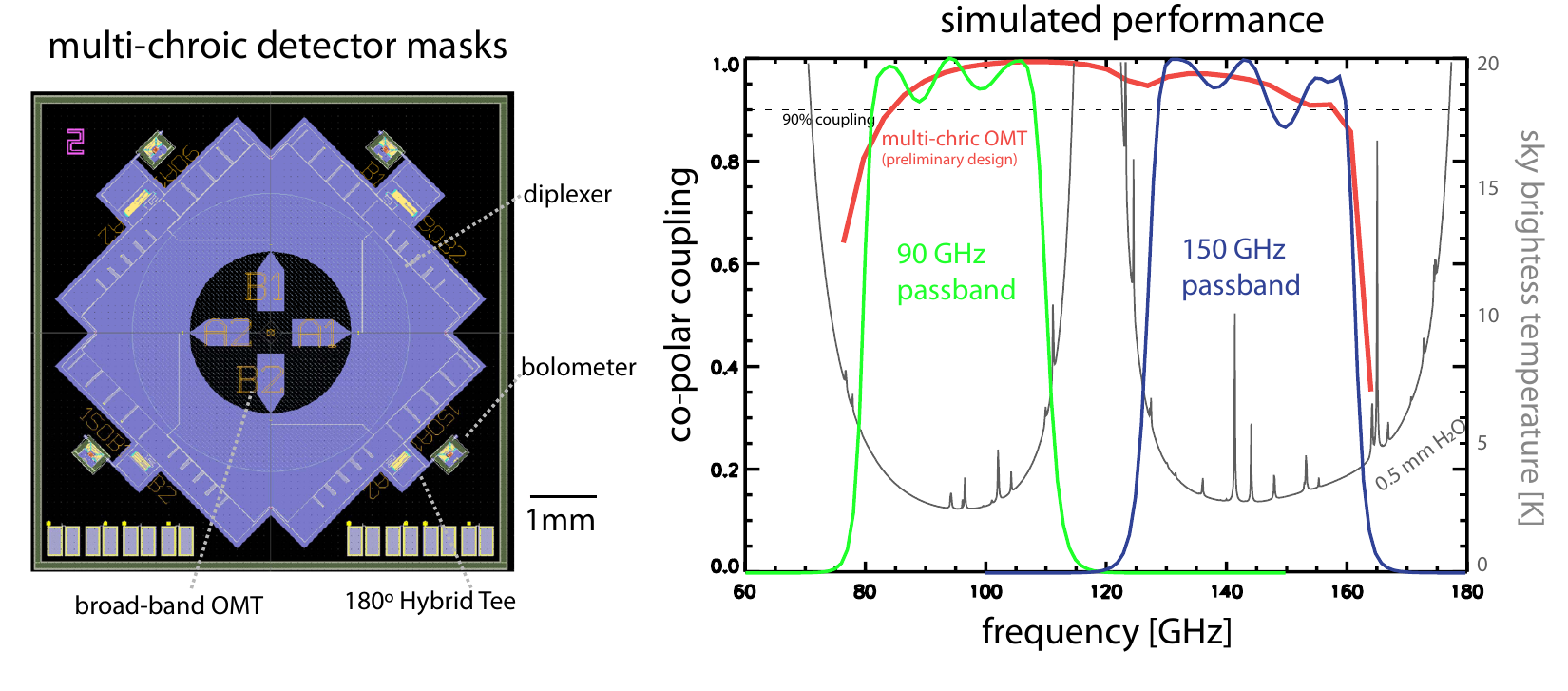}
\caption{{\it Left:} The layout for a multi-chroic pixel.  Light  in both frequency bands couples onto the OMT (center) before the fields from each fin are carried to the input diplexers, which split the radiation within each band into separate MS lines.  These lines pass through a cross-over before being carried into the inputs of $180^\circ$ hybrid tees.  Signals from unwanted waveguide modes are sent to the `sum' port (closest to the OMT) where they are terminated in gold resistors.  The desired TE11 mode is routed to a bolometer where it the signal is detected.      {\it Right:} The simulated performance of this system as a function of frequency.  The red curve shows the coupling of the OMT fins which is above 95\% averaged over either band.   The green and blue curves show the passband of the two diplexer outputs.  Cross-polarization (not shown) is below 1\%. The gray curve shows a simulation of the atmospheric emission at zenith at the ACT site, highlighting that these bands are well matched to the atmospheric transmission windows.  }
\label{fig:multichroicdet}
\end{figure}
 
With a few changes these detectors can be extended to octave (1:2) bandwidth. Octave bandwidth is sufficient to allow simultaneous operation at both the 90 and the 150 GHz frequency bands which lie on either side of the polarization foreground minimum, making these the most sensitive frequencies for CMB polarization studies.  


 Figure \ref{fig:multichroicdet} shows the detector layout for a multi-chroic 90/150 GHz pixel using a broader-band OMT based on Knchel and Mayer (1990)\cite{knochelandmayer}. The four OMT fins feed diplexers that split the 90 GHz and 150 GHz bands into separate MS lines. These diplexers are based on the resonant stub filters that are used in the single frequency detectors which have consistently matched simulations at the percent level. The outputs of the diplexers from opposing arms of the OMT are fed into the inputs of $180^\circ$ hybrid tee designed following \cite{Grimes:2007gs}, one for each frequency band and polarization state. This tee combines the signals from the two lines to produce a sum and a difference output, with the sum output terminated and the difference output routed to a bolometer island for detection. The hybrid tee is important because circular waveguide is multi-moded over octave bandwidth. The desired TE-11 mode couples to opposite probes with a $180^\circ$ relative phase shift (selected by the difference output of the hybrid tee), while all other modes over this bandwidth which couple to the probes are in phase (and thus terminated at the sum output). Simulations show the $180^\circ$ hybrid rejects 99.9\% of the unwanted modes with $> 99\%$ transmission of the desired TE-11 mode.  This establishes single mode performance over the full bandwidth (1:2.2), which is crucial to maintaining excellent polarization performance in this system.  In total, four bolometers are used to detect both linear polarizations in the two bands.


\begin{figure}[!t]
\includegraphics[width=3.5in]{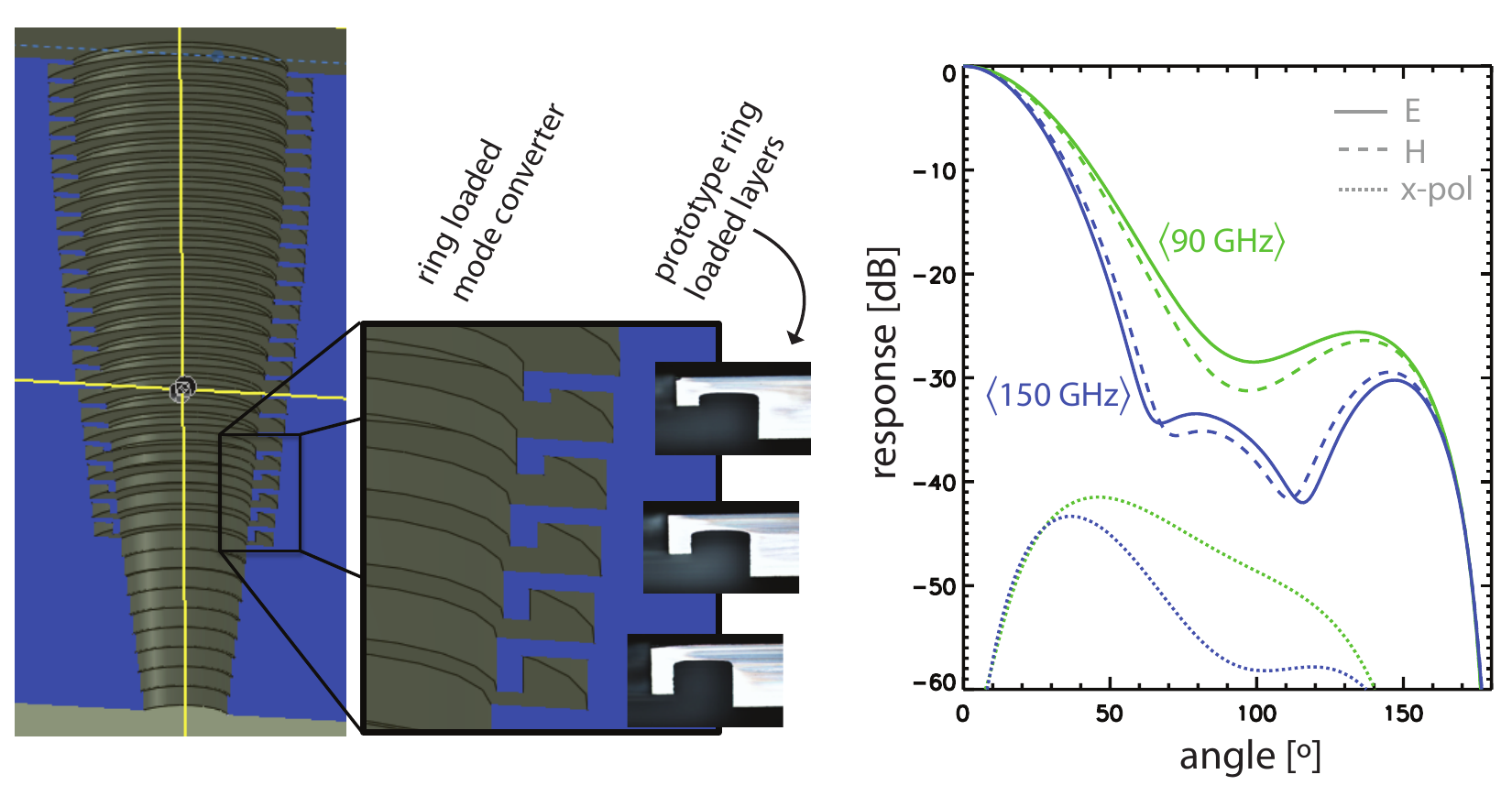}
\caption{\textit{Left:}  A drawing of a preliminary horn design incorporating ring loaded slots.  The zoom shows the geometry of the ring-loaded grooves more clearly.   Three photographs show prototype layers etched using a three layer mask and a deep reactive ion-etch (DRIE) machine. \textit{Right:}  The predicted band-averaged beam pattern in both the 90 and 150 GHz bands.   These patterns were constructed by simulating the beam pattern at 5 GHz increments and averaging these results within the predicted detector passband.  These simulations showed the input reflection to be below -20 dB and the cross-polarization below -30 dB across both bands. }
\label{fig:ringloaded}
\end{figure}

While standard corrugated horns do not offer sufficient bandwidth for these multi-chroic detectors, designs incorporating ring-loaded slots have demonstrated excellent performance for such applications \cite{ringloadedguide,EVLAhorn}. Figure \ref{fig:ringloaded} shows the design and simulations for such a horn. It achieves input reflection below 1\% and produces symmetric beams with peak cross-polarization below -30 dB over the entire band, from 80 to 160 GHz. This design is similar to a standard corrugated feed, but with the addition of ring-shaped cavities for the first 5 grooves.  Prototype ring loaded layers have been fabricated using  the NIST deep ion etch process \cite{DRIEprocess} with depths controlled to $\sim\!\!7$ micron accuracy, comparable to the tolerance of the standard electroforming technique. This silicon platelet array technology may prove to be the most economical method for producing large arrays of ring-loaded horns which are otherwise difficult to manufacture at mm-wave frequencies.

All detector components have been fully characterized using the same simulation tools (HFSS: www.ansoft.com and Sonnet: www.sonnetsoftware.com) used and extensively validated in the single frequency development. These simulations combined with estimates for dielectric losses predict an optical coupling efficiency of $> 70\%$,  polarization leakage below 1\% and excellent cross-polarization from the corrugated horn.			

\section{Multi-Chroic Detectors on ACTPol} 
		\label{sec:act}

	\begin{figure}[!t]
	\includegraphics[width=4.5in]{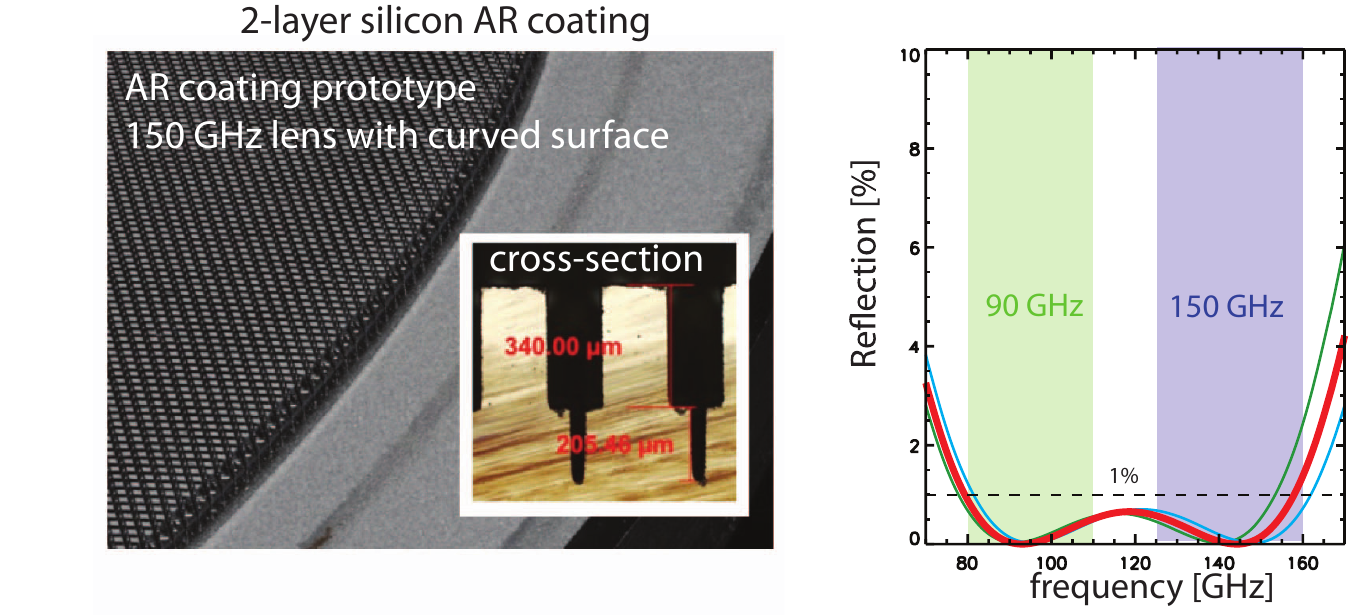}
	\caption{{\it Left:} Photographic images showing a top and side view of a two layer AR coating for silicon constructed by cutting sub-wavelength posts with a dicing saw on a curved lens surface. {\it Right:} The predicted reflection from this surface as a function of frequency between 80 and 160 GHz.  The red curve shows the reflection for this coating design.  The thin blue and green curves show the performance, including estimates for the measured machining errors. }
	\label{fig:ar}
	\end{figure}

As originally proposed, the ACTPol receiver was designed to use three single-frequency
lens-fed arrays of 512 polarimeters (see Niemack et al 2010~\cite{niemackactpol}).  
Each array is mounted in an `optics tube' which contains of a set of three silicon lenses that re-image light from the telescope onto the cryogenic detector array;  a cold Lyot stop to control illumination of the primary mirror;  and filters to block infrared radiation.  
The detectors are read out by time-domain multiplexing
(TDM) electronics provided by NIST \cite{dekorte03}, and are
controlled by room temperature Multi-Channel Electronics
\cite{battistelli08}, the same readout and control system
used for ACT. The ACTPol schedule
is to deploy a 150 GHz array in early 2012 followed by a second 150
GHz array and a third array in 2013.  We plan to deploy a third array consisting of 90/150 GHz multi-chroic detectors instead of the originally planned 220 GHz array.

\label{sec:detnoi}
The sensitivity of a single detector improves as the horn aperture grows larger.   For an array fabricated on a fixed diameter wafer (6$^{\prime\prime}$ silicon in this case), as the horn size grows larger, the number of detectors  decreases.  This leads to a particular number of horns (and horn size) that maximizes the array sensitivity.  Detailed calculations were performed, including a model of the telescope and atmospheric loading, consideration of bandwidth, and conservative estimates for the optical coupling.  This calculation, subject to the constraint of using the existing 1024 channel readout, revealed that 256 detectors was the optimal choice.  With this choice the array sensitivity is equivalent to an optimal 90 GHz array and  70\% of a 150 GHz-only array.  This  provides a significant boost in sensitivity compared with the originally planned 220 GHz array.  We note that the total array sensitivity could be further increased if we were not subjected to the constraint of using the existing readout.

The input to this array must pass through a set of silicon lenses.  We are developing a novel anti-reflective (AR) coating for the silicon lenses.  This `coating' consists of a two layer simulated dielectric layers constructed by removing grooves in two directions using a dicing saw to leave stepped pyramids on the lens surface. Figure \ref{fig:ar} shows a prototype of this coating on a curved surface and its simulated bandwidth.  This design offers extremely low loss and reduces reflections from each silicon surface below 1\% across a bandwidth of 1:2, which is sufficient for these multi-chroic pixels.  Metrology of the fabricated geometry show $\lesssim\!\!5$ micron deviations from the design, within the required tolerance.  

\section{Conclusions}
We have described our plans for extending the Truce single frequency horn coupled TES polarimeters to achieve multi-chroic performance.  We will initially develop detectors with sensitivity in the 90 and 150 GHz bands where CMB polarization foregrounds are lowest.  In the future, this architecture will be scaled to higher and lower frequencies to pursue a variety of science goals.  This approach retains the advantages of horn coupled polarimeters including symmetric beams, low cross-polarization, and frequency independent detector polarization angles while enhancing sensitivity and frequency coverage.  We plan to deploy an array of these detectors as part of the ACTPol project in 2013.  This will serve as a technology demonstration and enhance ACTPol's sensitivity to CMB polarization and gravitational lensing.

\label{sec:conc}

\begin{acknowledgements}
We would like to thank the members of the Truce collaboration for their many contributions to this architecture.
\end{acknowledgements}


\end{document}